\documentclass[manuscript,nonacm]{acmart}

\AtBeginDocument{%
  }

\usepackage{tabularx}
\usepackage{longtable}
\usepackage{booktabs}
\usepackage{xcolor}
\usepackage[table]{xcolor}
\usepackage[most]{tcolorbox}
\usepackage{fontawesome5}
\usepackage{soul}
\usepackage{array}
\usepackage{tikz}
\usepackage{graphicx}
\usetikzlibrary{decorations.pathreplacing, calc}

\definecolor{dimgray}{HTML}{5F6368}
\definecolor{dimgraylight}{HTML}{F1F2F3}

\colorlet{tech}{dimgray}
\colorlet{techlight}{dimgraylight}

\colorlet{integration}{dimgray}
\colorlet{integrationlight}{dimgraylight}

\colorlet{use}{dimgray}
\colorlet{uselight}{dimgraylight}

\definecolor{techhighlight}{HTML}{E4EFF7}
\definecolor{usehighlight}{HTML}{E6F2E5}
\definecolor{integrationhighlight}{HTML}{EEE7F5}

\newcommand{\techhl}[1]{{\sethlcolor{techhighlight}\hl{#1}}}
\newcommand{\usehl}[1]{{\sethlcolor{usehighlight}\hl{#1}}}
\newcommand{\integrationhl}[1]{{\sethlcolor{integrationhighlight}\hl{#1}}}

\newcommand{\techicon}{\textcolor{gray!65}{\faCog}}
\newcommand{\useicon}{\textcolor{gray!65}{\faUser}}
\newcommand{\integicon}{\textcolor{gray!65}{\faPlug}}

\newcommand{\xhdr}[1]{\vspace{1.5mm}\noindent\textbf{#1}}

\begin{document}


\title[How People in Decentralized Social Media Reason about the Appropriateness of Generative AI]{``Okay, I’ve Actually Softened My Take on This'': How People in Decentralized Social Media Reason about the Appropriateness of Generative AI}


\author{Romina Mahinpei}
\email{rmahinpei@princeton.edu}
\orcid{0000-0002-7500-5928}
\affiliation{%
  \institution{Princeton University}
  \country{United States}
}

\author{Manoel Horta Ribeiro}
\email{manoel@cs.princeton.edu}
\orcid{0000-0002-6159-9657}
\affiliation{%
  \institution{Princeton University}
  \country{United States}
}

\author{Andrés Monroy-Hernández}
\email{andresmh@princeton.edu}
\orcid{0000-0003-4889-9484}
\affiliation{%
  \institution{Princeton University}
  \country{United States}
}

\author{Sohyeon Hwang}
\email{sohyeon@princeton.edu}
\orcid{0000-0001-8415-7395}
\affiliation{%
  \institution{Princeton University}
  \country{United States}
}

\renewcommand{\shortauthors}{Mahinpei et al.}

\begin{abstract}
Generative AI (GenAI) is increasingly integrated into social media, raising questions about whether, where, and how it belongs. In decentralized social media (DSM), these decisions are distributed across users, developers, moderators, and administrators, making GenAI a collective governance challenge. At the same time, public discourse often flattens arguments to broad pro- or anti-AI positions that offer little insight into what people actually find (in)appropriate and why. Through 20 semi-structured interviews with people from Mastodon and Bluesky, structured around seven GenAI scenarios, we examine how people reason about GenAI's appropriateness in DSM. We find that participants drew conditional boundaries around particular GenAI configurations through distinct, salient, and weighted considerations spanning technology, integration, and use. We conceptualize this as boundary drawing and show how making such boundaries visible can support more grounded design, policy, and collective deliberation around GenAI in DSM.
\end{abstract}



\maketitle

\section{Introduction}
Generative artificial intelligence (GenAI) is increasingly embedded across social media~\cite{Pan25GenAISocialMedia,Moller26GenAISocialMedia}, shaping how content is created~\cite{CreationGenAI,VideoCreation}, 
information is curated and summarized~\cite{CurationGenAI1,CurationGenAI2,SummarizationGenAI}, 
and moderation is conducted~\cite{ImageModerationGenAI,Zhang26Moderation}. As these possibilities expand, platforms and users must decide whether, where, and how GenAI belongs in their shared spaces. For example, GenAI may help alleviate resource-constrained moderation work~\cite{Zhang26Moderation}, but its use may also reduce the perceived quality and authenticity of posted content~\cite{Moller26GenAISocialMedia}. These tensions shift the question from what GenAI \textit{can} do to what role users and communities \textit{want} it to play, and how to design systems or set policies accordingly.

Questions around the appropriateness of GenAI are particularly consequential in decentralized social media (DSM). Platforms like Mastodon and Bluesky have emerged as alternatives to centralized social media, motivated in part by the promise of decentralizing power, redistributing control, and giving users greater agency over their online experiences~\cite{Zulli20DsocialOverview,Abbing23DsocialOverview,Jhaver23DSocialOverview,Gehl23MastodonOverview}. Mastodon distributes governance across independently operated servers that establish their own rules, while Bluesky decentralizes components such as hosting, feed curation, and moderation so that individuals can choose service providers independently~\cite{Oshinowo25DsocialProtocols}. Nonetheless, both platforms distribute aspects of control across users, developers, moderators, and administrators~\cite{Legitime26GovernanceDsocial,Muralidharan26GovernanceDsocial,Hwang26GovernanceDsocial,Zhang25CommunityBlocklists}.
Whereas in centralized social media, the role of GenAI is largely determined by the companies running said platforms, in DSM, these decisions are distributed across multiple actors with potentially different interests, values, and responsibilities.

Likewise, decision-making about GenAI becomes distributed across many heterogeneous actors. In thinking about whether, how, and what kind of GenAI technologies are introduced, these actors will also need to anticipate how others in the ecosystem will respond. Misjudging those reactions can provoke substantial resistance and backlash. An illustrative case occurred at the start of 2026 when Bluesky developers released Attie, an LLM-powered tool designed to help users create custom feeds~\cite{BskyAttieUsesClaude}. Shortly after its release, Attie became the second-most-blocked account on Bluesky~\cite{BskyAttieSecondMostBlocked}, with users publicly expressing annoyance with AI features being rolled out\footnote{For example, one user post ended with the statement: ``WE WANT TO BE ABLE TO DISCOVER THINGS HERE ORGANICALLY WITH *HUMAN,* NOT MACHINE-ASSISTED LLM INTERACTIONS''}. At the same time, public discussions about Attie --- and about AI in DSM generally --- overwhelmingly describe attitudes toward GenAI simply as either ``pro-AI'' or ``anti-AI''~\cite{MastodonAntiAI,MastodonLeftistAI,MastodonUnavoidableAI,MastodonStronglyAntiAI,BskyBacklashAI,BskyDisgustAI}. For example, news coverage of Attie explained the backlash as being due to Bluesky having an anti-AI user base~\cite{BskyBacklashAI,BskyDisgustAI}, with the anti-AI characterization also being applied to the broader DSM ecosystem~\cite{MastodonAntiAI,MastodonStronglyAntiAI}.

However, attitudes about technologies are rarely neat dichotomies. Prior HCI research suggests that judgments about technologies, including GenAI, are often situated and conditional rather than reducible to stable positions for or against them~\cite{Lee18SituatedPerceptions,Kocielnik19SituatedPerceptions,Schroeder25LLMQualitative,Kyi25GenAICreativeWork,Anthis26AISenseMaking,Zhang26Moderation}. As researchers and developers seek to integrate GenAI features more and more into DSM --- such as feed curation~\cite{CurationGenAI1}, community-rule compliance~\cite{LaCava25GenAIComplianceDsocial}, or moderation support~\cite{Zhang26Moderation} --- understanding \textit{why} people are for or against such features is crucial to guide governance and decision-making in the DSM ecosystem at large. 
Designing systems and policies that can better align across heterogeneous stakeholders requires understanding how such judgments are structured: which considerations people draw on, how they shift across configurations, and how they combine to distinguish applications of GenAI that people consider appropriate from those they do not.

\xhdr{Present Work}.
Our work asks: \textit{How do people in decentralized social media reason about the appropriateness of GenAI within their shared spaces?} 
We explore this question through 20 semi-structured interviews with people active across Mastodon and Bluesky, presenting scenarios with different potential GenAI applications in DSM. Rather than expressing uniformly positive or negative positions, participants drew nuanced \textit{boundaries} around the appropriateness of GenAI: conditions under which particular GenAI configurations did or did not belong in their shared spaces. To do so, they reasoned through multiple considerations spanning three interconnected dimensions: the \textit{technology} itself and its implications; the \textit{integration} of GenAI, or how it is introduced within a shared space; and the \textit{use} of GenAI, or what it is being used for and how. Participants invoked different combinations of these dimensions, producing boundaries that varied both between and within participants across scenarios. 

Through our work, we aim to shift from broad characterizations of ``for'' or ``against'' attitudes to actionable insights that guide policy and system design. Our contributions are threefold:

\begin{itemize}
    \item \textbf{An empirical account of the considerations underlying GenAI judgments in DSM.}
    We identify the considerations people draw on when judging GenAI appropriateness in DSM and show that their salience and weight vary both across people and within the same person across scenarios.

    \item \textbf{Conceptualizing GenAI appropriateness as a process of drawing boundaries.}
    We show how differently weighed considerations spanning \textit{technology}, \textit{integration}, and \textit{use} combine to form conditional boundaries around GenAI, capturing the nuance that structures how people reason about GenAI in DSM.

    \item \textbf{Implications for governing GenAI in DSM.}
    We show how heterogeneous and sometimes only partially articulated boundaries complicate decision-making and identify explicit boundary articulation as a useful step toward collective deliberation around GenAI governance in DSM.
\end{itemize}

\section{Background}

Our work builds on three strands of research: the growing role of GenAI in social media and the governance questions it prompts (\S\ref{sec:bg_genai_sm}); decentralized social media and its distributed governance (\S\ref{sec:bg_dsm}); and research on technology appropriateness (\S\ref{sec:bg_tech_appropriateness}), which motivates examining how heterogeneous judgments around GenAI are structured.

\subsection{Emergence of Generative AI in Social Media}
\label{sec:bg_genai_sm}

GenAI introduces new possibilities for how people create, encounter, and govern content on social media. Researchers have already begun exploring how to leverage GenAI for content creation and conversational assistance~\cite{Moller26GenAISocialMedia,Radivojevic24GenAIDsocial}, summarizing online content and perspectives~\cite{SummarizationGenAI,RevisionsGenAI1}, evaluating or transforming content according to users' needs~\cite{ImageModerationGenAI,Ma25GenAIClassification}, and eliciting and implementing users' feed preferences~\cite{ElicitationGenAI,CurationGenAI2}. These capabilities can offer tangible benefits. For example, Rashed et al.~\cite{ImageModerationGenAI} used GenAI to transform potentially triggering images based on users' descriptions to allow users to avoid unwanted content while preserving its surrounding informational value, finding their approach increased participants' reported agency and safety. 

At the same time, introducing GenAI into online social spaces can create new concerns around privacy, authenticity, manipulation, bias, and the provenance of synthetic content~\cite{Chou26ChatbotsInSocialMedia,Pan25GenAISocialMedia}. Chou et al.~\cite{Chou26ChatbotsInSocialMedia}, for example, found that many group chat users were unaware of what information integrated chatbots could access and reported greater privacy concern after learning about that access. Such tensions extend longstanding questions about power and agency in social media. Prior scholarship and public-facing debate have discussed whether people recognize when algorithmic systems are shaping their experiences~\cite{Eslami15ReasoningInvisibleAlgorithms,Eslami16FolkTheories}, how they understand the consequences of those systems~\cite{Karizat21TikTokFolkTheories,DeVito17FolkTheories,Mayworm24ModerationFolkTheories}, and whether they have meaningful levers of change~\cite{Crawford16SMReportingTools,Ananny18AlgorithmicTransparency,Frey23VoiceOnlineCommunities}. Whereas earlier social media algorithms primarily shaped content curation and delivery, GenAI can participate in social media in a much broader range of ways. This potential for expansion raises the stakes of answering questions about when, where, and under what conditions GenAI involvement is appropriate or inappropriate.

As the capabilities of GenAI expand, platform operators and researchers have increasingly confronted questions about how they should be governed in social media. To date, the most visible responses have centered on \textit{AI-generated content}: platforms have introduced policies, disclosure requirements, and technical labels for synthetic media~\cite{Zahn24MetaAILabels,Malik25TiktokAILabels,Perez26AILables}. A recent analysis of 40 social media platforms found that more than two-thirds explicitly addressed AI-generated content in their governance policies~\cite{Gao26AIGCGovernance}. For instance, Meta has begun labeling what it predicts to be AI-generated images on Facebook, Instagram, and Threads in an effort to prevent the spread of misinformation and misleading content~\cite{Zahn24MetaAILabels}, with similar efforts seen in TikTok~\cite{Malik25TiktokAILabels} and LinkedIn~\cite{Perez26AILables}. At the same time, AI-generated content represents only one way GenAI can become embedded in social media. For other applications, such as GenAI summarizing or filtering content, questions around the appropriateness of GenAI remain considerably less investigated by researchers and practitioners alike.

\subsection{Governance in Decentralized Social Media}
\label{sec:bg_dsm}
Persistent concerns about the societal impacts of centralized, privately owned social media platforms (e.g., around privacy or algorithmic amplification ~\cite{Hinds20CambrdigeAnalyticaScandal,Massanari17RedditAlgGovernance}) have drawn attention to decentralized alternatives. Decentralized social media (DSM) encompasses platforms and protocols that distribute aspects of social media infrastructure and governance across multiple actors rather than concentrating them within a single provider~\cite{Abbing23DsocialOverview,Oshinowo25DsocialProtocols}. We focus on two prominent examples: Mastodon, built on the ActivityPub protocol as part of the broader Fediverse, and Bluesky, built on the AT Protocol. Both are microblogging platforms akin to Twitter/$\mathbb{X}$ that grew substantially following shifts in the social media landscape after Elon Musk's acquisition of Twitter in 2022~\cite{Jhaver23DSocialOverview,Ittefaq25DsocialMigration,Jeong24MastodonMigration,Cava23MastodonMigration,Quelle25BlueskyMigration}. Mastodon consists of independently operated servers/instances that establish their own rules and policies while federating with one another~\cite{Zulli20DsocialOverview,Gehl23MastodonOverview}, whereas Bluesky decentralizes components such as content hosting, feed curation, and moderation~\cite{Oshinowo25DsocialProtocols}.

Although they take different approaches to decentralization, both Mastodon and Bluesky redistribute forms of control that are typically concentrated within a single platform provider, creating opportunities for people in DSM to shape their spaces according to local needs and values~\cite{Hwang26GovernanceDsocial,Legitime26GovernanceDsocial}. Prior work has accordingly examined how communities and the various actors within DSM shape information flows~\cite{Hwang25OnlineCommunities}, federation and blocklist relationships~\cite{Zhang25CommunityBlocklists}, feed curation~\cite{Liu25DsocialFeedCuration}, and moderation rules~\cite{Muralidharan26GovernanceDsocial}. 
Returning to the question of GenAI, the redistribution of authority means a centralized provider does not automatically decide whether an application belongs; instead, different actors can participate in decisions about what is built, introduced, or used. In short, DSM is an empirical setting where people with different roles and values have salient, meaningful opportunities to negotiate the conditions under which GenAI is appropriate; this makes DSM a particularly valuable setting for studying how people reason about the appropriateness of GenAI.

Emerging research provides early examples of how questions around GenAI play out in particular parts of DSM. For example, existing work has explored using LLMs to curate personalized feeds~\cite{CurationGenAI1} and evaluate content against locally defined rules~\cite{LaCava25GenAIComplianceDsocial}. Closest to our work, Zhang et al.~\cite{Zhang26Moderation} examined how DSM moderators and administrators envisioned AI broadly supporting their moderation and governance work, identifying hypothetical roles such as providing contextual intelligence, facilitating cross-instance coordination, and reducing moderation burdens, alongside what they call ``governance boundaries'' (a set of normative values about where AI should be used). 
Where their work synthesizes distinct ideas for designing AI tools for moderation-related work, we turn our attention to how people across DSM may variably reason about \textit{generative} AI at large --- especially as GenAI presents a much broader range of applications that might enter the DSM ecosystem, in ways that are usually outside the control of one individual. We address this gap by investigating how users, developers, moderators, and administrators reason across multiple GenAI applications, focusing not only on where they place boundaries, but on the considerations and conditions through which those boundaries are constructed.

\subsection{Appropriateness of Technology and Generative AI}
\label{sec:bg_tech_appropriateness}

Research on technology acceptance has traditionally asked why people adopt or reject technologies, emphasizing factors such as perceived usefulness, ease of use, and social influence~\cite{Davis89TAM,Venkatesh03UTAUT}. These frameworks have also been extended to AI~\cite{Sapian26UTAUTExtension}, although researchers have argued that evaluations of AI can implicate additional considerations, including opacity, bias, fairness, and trust~\cite{Kelly23AIAcceptanceFactors,Liao22AITrust}. A complementary line of HCI research shifts attention from general acceptance toward \textit{social acceptability}, emphasizing that judgments about technology depend on how it fits the practices, expectations, relationships, and situations surrounding its use~\cite{Kocielnik19SituatedPerceptions,Lee18SituatedPerceptions,Uhde21SocialAcceptability}. Uhde et al.~\cite{Uhde21SocialAcceptability}, for instance, argue that frameworks such as the Technology Acceptance Model (TAM) are too rigid to capture how perceptions can vary across social situations, motivating greater attention to the context surrounding technology use. From this perspective, judgments about a technology may not reflect a stable position toward the technology as a whole: the same technology may be considered appropriate in one setting and inappropriate in another.

Emerging research on GenAI further highlights the conditional nature of such judgments. Across creative work, qualitative research, and social media, people distinguish among specific uses of GenAI based on how the technology is involved in an activity and the conditions surrounding its use~\cite{Schroeder25LLMQualitative,Kyi25GenAICreativeWork}. Kyi et al.~\cite{Kyi25GenAICreativeWork}, for example, interviewed creative workers and found that they evaluate GenAI in relation to where it enters the creative process and whether its development and use align with expectations around consent, credit, and compensation. Schroeder et al.~\cite{Schroeder25LLMQualitative} similarly interviewed qualitative researchers and found that they weigh possible LLM assistance against ethical concerns, model bias, participant interests, and privacy, resulting in different judgments across research tasks. In social media, reactions can likewise vary depending on whether GenAI assists or replaces human creation~\cite{Bruns24AIGC} or whether its involvement is disclosed~\cite{Jung25AILabelsPerceptions,Pawelczyk26AILabels}. Together, this work suggests that GenAI appropriateness is not simply a property of the technology itself, but depends on the particular context and conditions under which it is deployed.

Establishing that judgments are conditional, however, is only a first step for governance. Existing work often identifies whether a particular use is appropriate, or surfaces the constraints associated with a specific application, but tells us less about how these considerations relate to one another, how their salience shifts across applications, or how people can arrive at different boundaries even when considering the same application. These questions become especially important in DSM, where heterogeneous judgments may need to be made visible and negotiated among actors who share authority over their decentralized spaces. We therefore examine not only whether participants consider particular GenAI applications appropriate or not in DSM, but also the structure underlying those judgments: which considerations they draw on, how their salience shifts across configurations, and how they combine to shape the boundaries participants draw around GenAI.

\section{Methods}
We conducted semi-structured interviews (June-August 2026) with 20 individuals from Bluesky and Mastodon, each lasting 45-60 minutes over Zoom. All participants provided informed consent prior to participating, consented to the recording of their interviews, and received a \$30 USD digital gift card after completing the interview. Our Institutional Review Board reviewed and approved the study. 

\subsection{Participant Recruitment}
We recruited participants through public announcements on Bluesky and Mastodon as well as emails to individuals who had previously participated in related studies and were open to further contact. We received 49 responses to the screening survey linked in our public announcements and four responses through email outreach. 

We used information in the screening survey to verify that potential participants met our inclusion criteria: (1) having used either Bluesky or Mastodon for at least a year, to ensure familiarity with DSM, and (2) having at least 10 posts, followers, or followed accounts on either platform, to ensure some level of engagement. We followed a stratified sampling technique~\cite{Trost86StratifiedSampling}, seeking perspectives from people with different types of involvement within the ecosystem (users, developers, moderators, and administrators) as well as experience with Mastodon versus Bluesky, both of which may shape interactions with and sentiments around GenAI. Our goal was not to make systematic comparisons between role groups or platforms, but to ensure that our account of reasoning was informed by perspectives from across the ecosystem and could capture considerations that might arise from different forms of involvement or platforms.

We invited eligible respondents to participate in staggered waves, allowing ongoing interviews and preliminary analysis to inform subsequent recruitment. The first author conducted all interviews and wrote memos after each interview, documenting observations relevant to the research question. These memos were shared and discussed with the research team to inform subsequent data collection. We stopped recruitment after 20 interviews, when our interview memos and notes suggested that later interviews primarily elaborated on considerations already documented rather than raising substantively new ones.

Table~\ref{tab:participants} summarizes our interview participants. Our interview pool ultimately included 4 individuals from Mastodon, 8 individuals from Bluesky, and 8 individuals from both, with 12 participants having some extent of technical experience across these platforms and 8 participants having no relevant technical experience.

\begin{table}[t]
    \centering
    \small
    \caption{\textbf{Interview participants}. Our participants spanned five countries (USA, Canada, India, Germany, and the UK) and 11 Mastodon instances, and differed in their length of platform use and forms of technical and non-technical involvement in decentralized social media. A dash indicates the participant does not use that platform; $\bullet$ indicates the presence of that form of involvement for a participant, with involvement being aggregated across platforms. Operator refers to moderation or instance administration.}
    \Description{Participant characteristics for the 20 interview participants. Rows correspond to participants P1 through P20. Columns report years of use on Bluesky and Mastodon and whether each participant had experience as a user, developer, or operator. A dash indicates that the participant did not use that platform, and a filled circle indicates the presence of that form of involvement.}
    \label{tab:participants}
    \rowcolors{3}{}{gray!8}
    \begin{tabular}{@{}l c c >{\hspace{1em}}c c c@{}}
        \toprule
        \rowcolor{gray!20}
        & \multicolumn{2}{c}{\textbf{Years on}}
        & \multicolumn{3}{c@{}}{\textbf{Involvement}} \\
        \rowcolor{gray!20}
        \textbf{Participant} & \textsc{Bluesky} & \textsc{Mastodon}
        & \textsc{User} & \textsc{Developer} & \textsc{Operator} \\
        \midrule
        P1  & --  & 8  & $\bullet$ &           &           \\
        P2  & 3   & -- & $\bullet$ & $\bullet$ &           \\
        P3  & 3   & 4  & $\bullet$ & $\bullet$ &           \\
        P4  & 3   & 4  & $\bullet$ &           &           \\
        P5  & 3   & 2  & $\bullet$ & $\bullet$ &           \\
        P6  & 3   & -- & $\bullet$ &           &           \\
        P7  & 3   & 4  & $\bullet$ &           &           \\
        P8  & 3   & -- & $\bullet$ & $\bullet$ &           \\
        P9  & 1   & -- & $\bullet$ & $\bullet$ & $\bullet$ \\
        P10 & 2   & -- & $\bullet$ &           &           \\
        P11 & 1   & -- & $\bullet$ & $\bullet$ &           \\
        P12 & 3   & 5  & $\bullet$ & $\bullet$ & $\bullet$ \\
        P13 & 2   & 4  & $\bullet$ &           &           \\
        P14 & 3   & 10 & $\bullet$ & $\bullet$ &           \\
        P15 & 2   & -- & $\bullet$ &           &           \\
        P16 & 3   & -- & $\bullet$ &           &           \\
        P17 & --  & 9  & $\bullet$ &           & $\bullet$ \\
        P18 & --  & 8  & $\bullet$ & $\bullet$ & $\bullet$ \\
        P19 & --  & 4  & $\bullet$ & $\bullet$ & $\bullet$ \\
        P20 & 3   & 4  & $\bullet$ &           & $\bullet$ \\
        \bottomrule
    \end{tabular}
\end{table}

\subsection{Interview Protocol}
The first author conducted interviews using a semi-structured interview protocol developed and reviewed by all authors, and refined based on feedback from three pilot interviews conducted before recruitment. Interviews were recorded and automatically transcribed using Zoom, after which the first author reviewed each transcript in full against the audio recording to verify and correct the transcription. The interview protocol comprised the following four parts (see Appendix~\ref{app:interview_protocol} for the full protocol). 

\begin{enumerate}
    \item \textbf{Background and Prior Experience.} We began with background questions about participants' experiences with DSM, including how they joined and used the platforms and any prior experience using GenAI. 
    \item \textbf{Mental Models of AI and GenAI.} We then asked participants to describe their understanding of ``AI'' and ``GenAI'' before discussing specific applications, to ground the conversation in a shared understanding of GenAI and what technological objects they considered as such. 
    \item \textbf{Scenario Discussions.} We next presented participants with seven scenarios involving GenAI in DSM, inspired by applications seen in existing social media ecosystems or studied and prototyped in prior literature~\cite{CurationGenAI1,CurationGenAI2,ElicitationGenAI,SummarizationGenAI,CreationGenAI,RevisionsGenAI1,RevisionsGenAI2,ImageModerationGenAI}. 
    The scenarios spanned different roles GenAI might play in social media, from generating and modifying user content to summarizing and curating feeds. 
    Specifically, the scenarios involved GenAI: (1) generating a post from user-written bullet points, (2) generating an accompanying image for a post, (3) shortening a post to meet a character limit, (4) revising a post’s grammar and spelling, (5) summarizing posts in a feed, (6) curating a feed based on a user’s stated preferences, and (7) helping users elicit and articulate their preferences for content in a feed.
    For each scenario, we asked participants to think aloud about their thoughts, reactions, and the reasoning behind them as well as what changes (if any) might alter their views.

    \item \textbf{Broader Reflections on GenAI in DSM.} Finally, we asked participants to reflect more broadly on what role, if any, they wanted GenAI to play in the future of DSM, including applications they would welcome or want to avoid, and whether any of their views had changed or become clearer over the course of the interview.
\end{enumerate}

\subsection{Thematic Analysis}

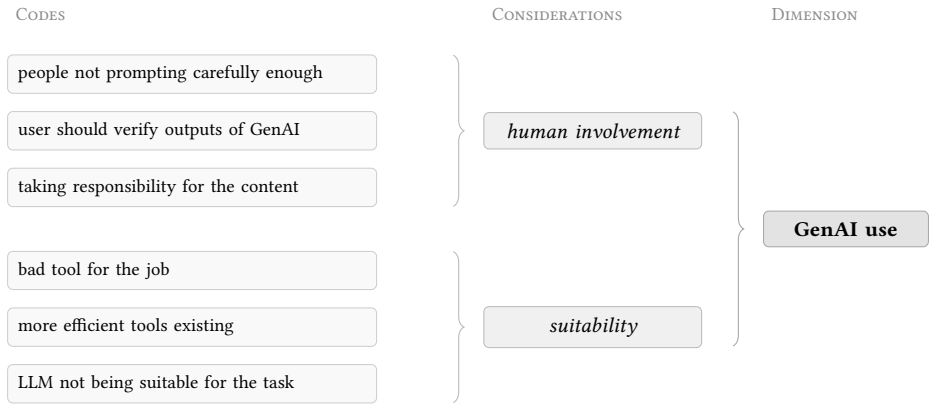
\begin{figure}[t]
\centering
\begin{tikzpicture}[
    code/.style={
        anchor=west,
        font=\footnotesize,
        draw=gray!40,
        fill=gray!5,
        very thin,
        rounded corners=2pt,
        inner sep=4pt,
        text width=4.6cm
    },
    mid/.style={
        anchor=west,
        font=\small\itshape,
        draw=gray!55,
        fill=gray!12,
        very thin,
        rounded corners=2pt,
        inner sep=4pt,
        text width=2.6cm,
        align=center
    },
    top/.style={
        anchor=west,
        font=\small\bfseries,
        draw=gray!70,
        fill=gray!22,
        very thin,
        rounded corners=2pt,
        inner sep=4pt,
        text width=1.9cm,
        align=center
    },
    hdr/.style={
        anchor=south west,
        font=\footnotesize\scshape,
        gray!90
    },
    midbrace/.style={
        decorate,
        decoration={brace, amplitude=4pt},
        gray!55
    },
    topbrace/.style={
        decorate,
        decoration={brace, amplitude=4pt},
        gray!70
    }
]

\node[hdr] at (0, 4.1)    {Codes};
\node[hdr] at (6.3, 4.1)  {Considerations};
\node[hdr] at (10.0, 4.1) {Dimension};

\node[code] (c1) at (0, 3.5)  {people not prompting carefully enough};
\node[code] (c2) at (0, 2.75) {user should verify outputs of GenAI};
\node[code] (c3) at (0, 2.0)  {taking responsibility for the content};
\node[code] (c4) at (0, 0.9)  {bad tool for the job};
\node[code] (c5) at (0, 0.15) {more efficient tools existing};
\node[code] (c6) at (0, -0.6) {LLM not being suitable for the task};

\draw[midbrace] (5.9, 3.75) -- (5.9, 1.75);
\draw[midbrace] (5.9, 1.15) -- (5.9, -0.85);

\node[mid] (m1) at (6.3, 2.75) {human involvement};
\node[mid] (m2) at (6.3, 0.15) {suitability};

\draw[topbrace] (9.6, 3.0) -- (9.6, -0.1);
\node[top] at (10.0, 1.45) {GenAI use};

\end{tikzpicture}
\caption{\textbf{Illustration of our thematic analysis process.} Initial codes were grouped into higher-level considerations, which were subsequently organized into the broader dimensions reported in our findings.}
\Description{An illustration of our thematic analysis process, showing how initial codes were grouped into higher-level considerations and then into a broader dimension for one example. Codes about prompting carefully, verifying GenAI outputs, and taking responsibility for content were grouped under human involvement. Codes about whether GenAI was a good tool for the job, whether more efficient tools existed, and whether an LLM was suitable for the task were grouped under suitability. These considerations were then organized under the broader use dimension.}
\label{fig:thematic_coding}
\end{figure}

Following data collection, we analyzed the interviews to characterize how participants reasoned about the appropriateness of GenAI in DSM, drawing on Braun and Clarke's principles of thematic analysis~\cite{Braun06ThematicAnalysis}. Our analysis began with inductive open coding, followed by the application and iterative refinement of a shared codebook. Two authors (the first and last authors) independently conducted line-by-line open coding of five interviews selected to capture variation in participants' platforms and forms of involvement. The two authors then met to discuss their codes, compare and reconcile them, and identify recurring concepts. Based on this process, the authors developed a shared codebook, which was subsequently reviewed with the whole research team (see Appendix~\ref{app:codebook} for the codebook).

The first author then applied the codebook to the remaining interviews. During this process, regular team meetings facilitated discussion of the data, with the first author discussing any ambiguities or data that suggested potential revisions or additions to the codebook. This resulted in revisions to three existing codes, regarding human involvement, suitability of GenAI for a task, and testing and iteration, as well as the addition of two codes capturing concerns about GenAI's opacity and bias and expectations that GenAI features keep users' and communities' interests at the forefront (see Appendix~\ref{app:codebook} for the full codebook and these changes). After coding was complete, the first and last authors examined relationships among codes across participants and scenarios, synthesizing related codes into higher-level themes and patterns capturing participants' reasoning about GenAI's appropriateness in DSM (see Figure~\ref{fig:thematic_coding} for a visual example). These themes and patterns were then discussed and iteratively revised with the whole research team.


\section{Findings}

\begin{figure*}[t]
\centering
\resizebox{\textwidth}{!}{%
\begin{tikzpicture}[
    title/.style={
        font=\bfseries\Large,
        align=center
    },
    subtitle/.style={
        font=\itshape\small,
        align=center
    },
    group/.style={
        anchor=west,
        font=\bfseries\small,
        align=left
    },
    item/.style={
        anchor=west,
        font=\small,
        align=left
    },
    icon/.style={
        circle,
        minimum size=0.72cm,
        inner sep=0pt,
        font=\large
    }
]

\def\colW{5.35}
\def\gap{0.14}
\def\headH{1.35}
\def\bodyH{5.55}

\pgfmathsetmacro{\xA}{0}
\pgfmathsetmacro{\xB}{\colW+\gap}
\pgfmathsetmacro{\xC}{2*(\colW+\gap)}

\def\iconX{0.62}
\def\textX{1.18}

\def\yTop{4.80}
\def\yMid{3.00}
\def\yBot{1.22}

\def\groupOffset{0.18}
\def\itemOneOffset{-0.22}
\def\itemTwoOffset{-0.56}

\newcommand{\dimensionitem}[8]{%
    \node[icon, fill=#3, text=#4] at (#1+\iconX,#2) {#5};
    \node[group, text=#4] at (#1+\textX,#2+\groupOffset) {#6};
    \node[item, text=#4]  at (#1+\textX,#2+\itemOneOffset) {#7};
    \node[item, text=#4]  at (#1+\textX,#2+\itemTwoOffset) {#8};
}


\fill[techlight]
    (\xA,\bodyH) rectangle ++(\colW,\headH);

\draw[tech, line width=0.6pt]
    (\xA,0) rectangle ++(\colW,\headH+\bodyH);

\draw[tech, line width=0.6pt]
    (\xA,\bodyH) -- ++(\colW,0);

\fill[integrationlight]
    (\xB,\bodyH) rectangle ++(\colW,\headH);

\draw[integration, line width=0.6pt]
    (\xB,0) rectangle ++(\colW,\headH+\bodyH);

\draw[integration, line width=0.6pt]
    (\xB,\bodyH) -- ++(\colW,0);

\fill[uselight]
    (\xC,\bodyH) rectangle ++(\colW,\headH);

\draw[use, line width=0.6pt]
    (\xC,0) rectangle ++(\colW,\headH+\bodyH);

\draw[use, line width=0.6pt]
    (\xC,\bodyH) -- ++(\colW,0);


\node[title, text=tech]
    at (\xA+0.5*\colW,\bodyH+0.90)
    {Technology};

\node[subtitle, text=tech]
    at (\xA+0.5*\colW,\bodyH+0.42)
    {What GenAI is and what it entails};

\node[title, text=integration]
    at (\xB+0.5*\colW,\bodyH+0.90)
    {Integration};

\node[subtitle, text=integration]
    at (\xB+0.5*\colW,\bodyH+0.42)
    {How GenAI is introduced into shared spaces};

\node[title, text=use]
    at (\xC+0.5*\colW,\bodyH+0.90)
    {Use};

\node[subtitle, text=use]
    at (\xC+0.5*\colW,\bodyH+0.42)
    {What GenAI is used for and how};


\dimensionitem{\xA}{\yTop}{techlight}{tech}{\faGlobeAmericas}
    {Societal Costs \& Impacts}
    {Environmental costs}
    {Training data \& consent}

\dimensionitem{\xA}{\yMid}{techlight}{tech}{\faServer}
    {Who Provides \& Controls It}
    {Local vs.\ cloud hosting}
    {Data privacy}

\dimensionitem{\xA}{\yBot}{techlight}{tech}{\faCheck}
    {Trust \& Reliability}
    {Bias \& opacity}
    {Hallucinations \& errors}


\dimensionitem{\xB}{\yTop}{integrationlight}{integration}{\faUsers}
    {Community Alignment}
    {User interests}
    {Community needs}

\dimensionitem{\xB}{\yMid}{integrationlight}{integration}{\faSlidersH}
    {Control \& Agency}
    {Opt-in / opt-out}
    {Customizable}

\dimensionitem{\xB}{\yBot}{integrationlight}{integration}{\faEye}
    {Transparency}
    {Disclosure of use}
    {Visibility into how it works}


\dimensionitem{\xC}{\yTop}{uselight}{use}{\faPen}
    {What GenAI Produces}
    {Assist vs.\ offload}
    {Degree of transformation}

\dimensionitem{\xC}{\yMid}{uselight}{use}{\faUser}
    {Role of Human}
    {Human in control}
    {Preserve authenticity}

\dimensionitem{\xC}{\yBot}{uselight}{use}{\faStar}
    {Suitability \& Value}
    {Capable of the task}
    {Needed for the task}

\end{tikzpicture}%
}

\caption{\textbf{Dimensions participants considered when reasoning about GenAI.} When reasoning about the appropriateness of GenAI in decentralized social media, participants considered GenAI as a technology itself, how it was integrated into shared spaces, and how it was used. We show example considerations within each dimension.}
\Description{Three-column diagram showing the dimensions participants considered when reasoning about GenAI in DSM along with example considerations. The technology dimension includes societal costs and impacts, who provides and controls GenAI, and trust and reliability. The integration dimension includes community alignment, control and agency, and transparency. The Use dimension includes what GenAI produces, the role of the human, and suitability and value.}
\label{fig:boundary_dimensions}
\end{figure*}

When presented with concrete scenarios involving GenAI in DSM, participants rarely treated its appropriateness as a simple question of whether the technology was ``good'' or ``bad.'' Instead, they reasoned about it by \textit{drawing boundaries}, specifying nuanced conditions under which particular GenAI scenarios could be appropriate or inappropriate. These boundaries reflected considerations that clustered around the three dimensions shown in Figure~\ref{fig:boundary_dimensions}, and reported in the first three findings sections below: the \textit{technology} itself and its implications (\S\ref{sec:findings_tech}); the \textit{integration} of GenAI, or how it is introduced into shared spaces (\S\ref{sec:findings_integration}); and the \textit{use} of GenAI, or what it is being used for and how (\S\ref{sec:findings_use}). 

Participants' reasoning was also shaped by what they valued about DSM: alternatives to centralized infrastructure, greater individual and collective agency, and spaces oriented around authentic interaction between people. These values surfaced differently across the three dimensions, rather than constituting a separate set of considerations, and did not produce a common position on where GenAI belonged. Importantly, as we describe in \S\ref{sec:findings_drawing_boundaries}, participants did not always invoke every dimension when evaluating a scenario, nor did the considerations they reasoned through carry equal weight in where they drew their boundaries.

\subsection{Technology: What GenAI Is and What It Entails}
\label{sec:findings_tech}

Participants' boundaries were shaped in part by how they perceived GenAI as a technology and its broader implications. Considerations included the societal costs associated with developing and operating GenAI, the reliability and biases of model outputs, and the infrastructural and institutional arrangements through which GenAI was provided. These latter considerations often mirrored participants' attitudes toward centralized infrastructures of mainstream social media and why they turned to decentralized alternatives in the first place. 

\subsubsection{\textbf{Participants considered the broader societal costs and impacts of GenAI}}
\label{sec:findings_tech_costs}
Participants frequently referenced the societal costs associated with GenAI, including the environmental and resource costs of operating models and using others' work as training data without permission. P7, for example, noted:
\begin{quote}
    \textit{``Generative AI and all the hype around it, just has \textbf{a lot of issues around, labor, copyright, and also environmental justice issues}. And then I just feel like in general generative AI kind of \textbf{heightens all the things wrong with society} because of the way it captures capital and then doesn't distribute it to everyone, like in the hands of a few.''} --\textbf{P7}
\end{quote}
For P7, GenAI embodied societal problems of inequality, with its benefits concentrated ``in the hands of a few.'' P17 similarly discussed GenAI in terms of costs and benefits, describing the societal costs of GenAI as ``pretty indisputable'' while finding its benefits unclear. 

Such concerns, however, did not necessarily produce immutable boundaries. P17 noted that reducing environmental consumption or developing some ``equitable way of getting consent and providing compensation for the use of training data'' could indeed ``change some of [their] opinions on it.'' P12 likewise raised several societal costs across scenarios, including environmental concerns and the ``authorship issue'' arising from GenAI being ``trained on big bodies of work.'' At the same time, they acknowledged that some of these concerns could become less salient as the technology changed, pointing, for example, to people ``running locally hosted models.'' For P12 and P17, their boundaries thus depended partly on how the technology was developed, and changing those conditions could change what they considered appropriate. 

Importantly, raising a potential societal cost did not necessarily define where a participant drew their boundary right away. P15, for example, was ``not immune to the environmental concerns,'' but explained that they were ``not particularly forefront'' for them. Instead, other issues like whether they could ``trust [their] information in some of [infrastructure]'' controlled by large corporations carried greater weight in shaping their boundary.

\subsubsection{\textbf{Participants considered who provided and controlled GenAI}}
\label{sec:findings_tech_control}

Participants considered the infrastructural and institutional arrangements through which GenAI was provided and accessed. Some participants were particularly wary of centralized GenAI infrastructure, which they saw as reproducing the concentration of power they associated with centralized social media. P5, for example, worried about ``having a couple of really large corporations building all of the infrastructure,'' which they believed ``could be harmful in the same way as those [...] same companies were building centralized social media.''

Concerns about who provided and owned GenAI were also tied to concerns about privacy. P13 mentioned that if the AI is ``hosted on Google, Microsoft, Meta servers'' that would draw ``the line in the sand because if it belongs to a monopoly, it's privacy zuckering.''\footnote{``Zuckering'' is an informal reference to Mark Zuckerberg, used here to evoke privacy concerns associated with large technology platforms' collection and control of user data.} In response to such sentiments, participants described arrangements that shifted ownership toward users or communities more favorably. P11 argued that users should ``own everything, the data and the AI, everything needs to be on there at their home, on their equipment, not in the cloud, not with private companies'' while P9 imagined using ``decentralized infrastructure to create AI that is more communally owned or community controlled'' such as by ``creating some sort of AI cooperative where everybody is pooling their compute together to be able to use AI in the ways that they need to without necessarily creating a reliance on big, major AI players.'' 

Similarly, P18 distinguished between ``a low-power local LLM'' and systems that relied on services such as ChatGPT or Claude (i.e., through an API), particularly when private content such as direct messages would then be shared with an external provider (such as OpenAI or Anthropic). They preferred the local LLM because it constrained where the model operated and who could access the data. For such participants, decentralization was thus relevant not only to the social platform itself but also to the infrastructure through which GenAI was being provided. 

\subsubsection{\textbf{Participants considered whether GenAI could be trusted}}
\label{sec:findings_tech_trust}
Participants also considered GenAI's reliability, pointing to past documented failures in model outputs. P6 pointed to GenAI's ``hallucination problem,'' while P20 recalled models producing ``interesting sounding sources that weren't right.'' Others foregrounded bias: P2 noted that ``the models themselves can be very biased,'' while P9 discussed how ``a lot of these AI models historically [and today] produced biased outcomes, specifically against communities of color.'' Some participants also considered how such errors or biases might be difficult for users to recognize. P14, for example, argued that ``transmitting bias'' was too simplistic of a description because GenAI tools were opaque, making it hard to know ``the degree to which an LLM is intentionally or inadvertently leveraging things.''

How much these reliability concerns mattered, however, depended on what GenAI was being used to do. P14 argued that using them to produce text that would be viewed as ``meaningful and truthful'' would need to be ``handled with care because of the kinds of mistakes [LLMs] make and the limitations that we know that they have.'' In contrast, they viewed creating a short one- or two-sentence summary of a post as ``probably fine because there's not a lot of room to get it wrong.'' Similarly, P13 felt that image generation required greater trust because the model could introduce ``a bunch of additional things you did not intend to say.'' They were more comfortable with an LLM shortening a text, where the user could ``try it and see'' whether it adequately reflects their intent.

\subsection{Integration: How GenAI is Introduced into Shared Spaces}
\label{sec:findings_integration}
Participants' boundaries were also shaped, in part, by how GenAI was deployed and introduced into their shared spaces. Given the context of DSM, where decentralized governance is a major principle, participants considered whether integrating GenAI into the DSM ecosystem aligned with the community norms and values, how much control individuals retained over any GenAI features, and whether GenAI's involvement was sufficiently transparent to all. 

\subsubsection{\textbf{Participants considered whether GenAI aligned with community norms and values}}
\label{sec:findings_integration_aligned}

Participants considered whether introducing GenAI aligned with the norms, values, and priorities of the spaces into which it was being integrated. These considerations focused less on whether GenAI could provide a helpful capability in isolation and more on whether introducing that capability was consistent with what community members wanted from their shared spaces. 
P8, for example, described themselves as being ``generally against having trendy things or a thing that I use just because it's trendy,'' arguing that people should more often ask ``why they are a trend, why they are a thing, and why they're useful.'' P14 similarly cautioned against being ``fooled into accepting things that are superficially surprising and remarkable,'' highlighting how novelty or technical capability did not itself establish that GenAI was appropriate. Failing to look more carefully at the concrete wants, goals, and needs of individuals within DSM may explain why the release of GenAI tools met backlash and resistance in those spaces. P17 noted: 

\begin{quote}
     \textit{``I think the main issue that I keep running into is seeing developers create [GenAI] tools that are \textbf{counter to the needs and the wants of people on the Fediverse}, and not really taking their views and their beliefs into account. So I think the main thing would be like, \textbf{look at the actual people on the Fediverse and what they care about} and take that into account when deciding what to build or what not to build.''}  \textbf{--P17}
\end{quote}

In this sense, alignment depended less on the presence of GenAI itself and more on whether its use preserved or supported what participants valued about their shared spaces. For example, participants who valued DSM as a space for authentic expression and interaction, like P7 and P8, were receptive to scenarios that made participation more accessible (e.g., translation, alt-text generation) while remaining more skeptical of those that substituted generated content for that interaction. Similarly, participants like P10, who cared about the safety of all involved in DSM, suggested that GenAI could ``improve the user experience'' when used to support moderation, both by helping moderators manage harmful content and by making the platform safer for participants. For such participants, GenAI could be seen as aligned with a space when it reinforced valued aspects of participation rather than displacing them.

\subsubsection{\textbf{Participants considered how much control individuals retained}}
\label{sec:findings_integration_control}
Participants further reasoned about whether people retained ``choice and agency over how they want these very important communication infrastructures to be run'' (P7). This included being able to choose whether to use a GenAI feature or to opt out of an integration. P5 saw making GenAI ``the only way that you could interact with the network from the application'' as dangerous, worrying that doing so would obscure where information came from and prevent users from directly accessing the underlying content. P8 similarly emphasized how simplifying the user experience should not mean ``taking away [user] agency entirely'' and favored giving users ``the option to turn it off,'' referring to unwanted GenAI features or integrations. Here, the risks that a GenAI feature would undermine human agency (for example, by constraining how users could access or interact with the platform) made it important for people to at least have control by being able to ``turn it off.''

Participants' boundaries also involved considerations about \textit{who} had the power to deploy something like a GenAI feature and why they might do so. P16 raised these questions when considering GenAI-mediated feed curation:
\begin{quote}
    \textit{``Well, so I think I would go back again to \textbf{how much control and who has it}, right? If you're letting the users say what they want to see [in a feed], and then letting them change that in a convenient conversational format [referencing GenAI feature], I think that's \textbf{just giving folks more knobs to twiddle their own personalized algorithm}, which strikes me as being mostly good. But like I say, if you're \textbf{building all the knobs and giving it to someone else with the profit motive there}, it's probably less good.''} -- \textbf{P16}
\end{quote}
For P16, the same broad scenario of GenAI curating a feed could therefore fall on different sides of their boundary depending on where control was located. Giving individuals additional ways to articulate and revise their own preferences around the hypothetical GenAI feature extended their agency; deploying similar features to give another actor greater influence over what people saw raised different concerns (e.g., selling user data). In short, GenAI was appropriate to the extent that it empowered the idea of user agency participants associated with decentralized systems.

\subsubsection{\textbf{Participants considered whether and how GenAI's role would be made transparent}}
\label{sec:findings_integration_transparency}

Participants drew boundaries around whether people could recognize GenAI's involvement in the interactions they had online with a clear understanding of the role it played. What counted as sufficient transparency, however, depended on what GenAI was doing. For content generation, participants often focused on \textit{disclosure}. P1 suggested that a profile could state that ``posts may contain AI-generated images'' or that generated images could be identified through their alt text. P9 similarly wanted generated images to be ``clearly labeled'' so that people could distinguish them from non-generated images. P4 extended this expectation across both image and text generation, arguing that people should disclose GenAI's involvement across both cases: ``I think you should say, we used AI in this way.''

As GenAI became more deeply integrated into the platform experience across the scenarios (e.g., summarizing or curating a feed), simply knowing that ``AI was used'' could be insufficient. In such cases, participants often wanted greater visibility into how GenAI shaped their experiences. P12, for example, valued giving people ``more control and more transparency into how [GenAI-based] recommendation algorithms work,'' while P19 described a ``black box'' GenAI system in which users neither know what's going on nor have control as ``a big problem in the social media space.'' Transparency could thus require different forms of visibility and levels of explanation depending on the integration: it could require disclosing GenAI's role in generated content in one case, and understanding how it influenced what people encountered in another case.

\subsection{Use: What GenAI Is Used For and How}
\label{sec:findings_use}
In addition to GenAI as a technology and its integration into shared spaces, we found that participants' boundaries were shaped by what GenAI was used to do and how. Participants considered what GenAI produced, its role (relative to the user), and its suitability. These considerations were often grounded in participants' understanding of social media as a space for people to express themselves and interact with one another authentically. Uses that supported people's ability to participate in this way could therefore be viewed differently from uses that appeared to substitute generated content or to automate interaction for that participation. 

\subsubsection{\textbf{Participants considered what GenAI was being asked to produce}}
\label{sec:findings_use_tranformation}
Participants distinguished between uses of GenAI that could deviate more versus less from a person’s original input. These distinctions were particularly salient when comparing scenarios in which GenAI generated a post from bullet points or produced an image with those in which it shortened a post or corrected grammar. P2, for example, described shortening and grammar correction as uses in which one's ``intended message will almost certainly be present.'' By contrast, generating text or an image had much more potential to move farther and dilute one's initial input, ultimately becoming a ``betrayal of [one's] meaning space.'' P3 similarly described image generation, where one is ``creating one medium from another,'' as the ``most transformative'' and therefore more likely to lose the person’s intended message. 

Greater transformation, however, did not necessarily make a use inappropriate. Participants weighed concerns about preserving original intentions or contributions against what kinds of interaction GenAI's involvement enabled. P1 wrestled with this tension when considering expanding bullet points into a post with a hypothetical GenAI feature:
\begin{quote}
    \textit{``I feel like there is a \textbf{benefit to it}, also. There are people who have been \textbf{afraid to post their thoughts} because blogging was difficult for them or, you know, social media was difficult for them, and now this tool is \textbf{giving them [a] voice}. But then, at the same time, \textbf{is it their voice?} It's a thought that the LLM expanded into a whole post, so what are we even looking at at that point?''} --\textbf{P1}
\end{quote}
Their uncertainty captures how these considerations about what GenAI produced could pull participants in different directions: the same transformation that raised questions about whether the resulting post still represented a person's voice could also enable someone to express something they otherwise might not have shared.

\subsubsection{\textbf{Participants considered what role remained for the person}}
\label{sec:findings_use_role}

Participants’ boundaries also depended on what role remained for the person when GenAI was involved. This often meant staying actively involved in the task rather than having GenAI perform it fully. P4 described ``a line between [being involved] and having [GenAI] just do everything for you,'' distinguishing full automation from uses where AI improved content originally produced by the user. P9 also drew a line between ``using LLMs to replace the process of generating ideas versus helping you after the fact,'' viewing the latter more favorably.

For others, what mattered was not simply how much GenAI did, but whether the person maintained meaningful control. P13, for example, described GenAI as ``a tool in [one's] arsenal'' when used as a light assistant, while P16 emphasized the importance of ``how much mastery [one is] showing over the tool.'' For these participants, whether the use was appropriate depended on how much the person remained actively in control. P8 similarly contrasted having GenAI ``rewrite this for me and then post it'' against approaches where a person could ``take piece by piece what exactly [they] like or don't like about the rewritten post.''

Participants also varied in what role they felt people should retain. P3 cared less about whether the words originated from a person than about ``how much ownership [someone] take[s] over the final output.'' Others focused more on the distinction between GenAI supporting and GenAI undermining human-to-human interaction, which they saw as the whole point of social media. P1 worried that GenAI did not have ``enough context to capture [their] voice,'' imagining a potential future in which social media became ``LLMs talking to each other.'' P20 similarly emphasized that ``the point is people talking to each other'' and was more receptive to GenAI assistance so long as the resulting interaction remained fundamentally human-to-human. For such participants, their reasoning reflected the view that DSM is not simply an infrastructure for distributing content, but a space whose value depends on authentic interaction between people.

\subsubsection{\textbf{Participants considered whether GenAI was suitable and added meaningful value}}
\label{sec:findings_use_value}
GenAI's ability to potentially do a task was not necessarily sufficient justification for using it. Participants often asked whether it addressed a meaningful need. 
P6, for instance, questioned the value of generating images for ordinary social media posts when people could instead share existing images or their own creations. P19 similarly questioned whether GenAI was the appropriate means for tasks such as correcting grammar when non-GenAI grammar checks already exist, which they personally found to be better than GenAI at ``pointing out those mistakes,'' referring to it as a case of GenAI being ``bolted onto [a task] that [it] is not really made for and not especially good at.'' For both P6 and P19, the existence of an older --- and usually simpler --- alternative seemed to render GenAI superfluous; the alternative was considered more suitable for the task at hand. In contrast, GenAI became meaningful for tasks such as translation (P7) or generating alt-text for images (P8). Participants distinguished these use cases because they showed scenarios where GenAI helped overcome a constraint (e.g., translation requires language skills and can help non-native English speakers; good alt-text takes time and/or can be hard for people who cannot type extensively) that ultimately made content more accessible.

The value participants attributed to a particular GenAI capability could also depend on the setting in which it appeared. P18, for example, questioned the value of summarizing ordinary social interactions, where reading and engaging with other people's posts was itself part of what they valued about social media, but saw a clearer role for summarization as a first pass for moderators facing substantial volumes of content. They specifically imagined GenAI helping moderators sort content into ``buckets'' or flagging potentially ``extremely distressing content'' before a human reviewed it, reducing moderators' burden while still keeping ``a human in the middle.''

\subsection{How Considerations Come Together to Form Boundaries}
\label{sec:findings_drawing_boundaries}
Across interviews, we found that boundaries around GenAI were multifaceted and complex, as also articulated by P11:
\begin{quote}
    \textit{``You’ve got boundaries yourself personally, spiritually, morally, physically, right? As a person, as a business, people all have these boundaries, so \textbf{where are your boundaries}? And then \textbf{AI needs to fit within those}. That’s the key.’’} --\textbf{P11} 
\end{quote}
Although we described the three dimensions separately, participants often considered them in combination when reasoning about GenAI. At the same time, they did not necessarily invoke all three dimensions when evaluating a given scenario. Some boundaries were shaped primarily by a consideration within one dimension, while others emerged from weighing several considerations within or across dimensions. Moreover, recognizing a consideration did not mean that it ultimately defined the boundary: participants could give it relatively little weight, weigh it against other considerations, or identify conditions that sufficiently addressed it. We observed three ways in which this process produced conditional and heterogeneous boundaries: boundaries differed across participants considering the same scenario (\S\ref{sec:findings_boundaries_diff_participants}), varied within participants across scenarios (\S\ref{sec:findings_boundaries_same_participants}), and could become further articulated as participants reasoned through concrete cases (\S\ref{sec:findings_boundaries_eliciation}).

\subsubsection{\textbf{The same scenario produced different boundaries across participants}}
\label{sec:findings_boundaries_diff_participants}

\begin{table}[t]
    \centering
    \footnotesize
    \caption{\textbf{Variation in participants' reasoning across GenAI scenarios.} 
    Based on our thematic coding, $\blacksquare$ indicates that the participant foregrounded that dimension when reasoning about the scenario; $\square$ indicates they did not. The selected participants and scenarios are illustrative rather than representative of the prevalence of particular reasoning patterns.}
    \Description{Matrix showing which of the technology, integration, and use dimensions were foregrounded in the reasoning of participants P4, P9, and P12 across three GenAI scenarios: writing a post, generating an image, and summarizing a feed. A filled square indicates that a dimension was foregrounded in the participant's reasoning for that scenario, while an empty square indicates that it was not. The pattern illustrates variation both across participants considering the same scenario and within participants across different scenarios.}
    \label{tab:reasoning}
    \renewcommand{\arraystretch}{1.25}
    \setlength{\tabcolsep}{4pt}

    \begin{tabular}{@{}l ccc ccc ccc@{}}
        \toprule
        & \multicolumn{3}{c}{\textbf{GenAI Writing a post}}
        & \multicolumn{3}{c}{\textbf{GenAI Generating an image}}
        & \multicolumn{3}{c}{\textbf{GenAI Summarizing a feed}} \\
        
        \cmidrule(lr){2-4}
        \cmidrule(lr){5-7}
        \cmidrule(lr){8-10}
        
        \textbf{P\#}
        & \techicon~\textbf{Technology}
        & \integicon~\textbf{Integration}
        & \useicon~\textbf{Use}
        & \techicon~\textbf{Technology}
        & \integicon~\textbf{Integration}
        & \useicon~\textbf{Use}
        & \techicon~\textbf{Technology}
        & \integicon~\textbf{Integration}
        & \useicon~\textbf{Use} \\
        \midrule

        \textbf{P4}
        & $\square$ & $\square$ & $\blacksquare$
        & $\square$ & $\blacksquare$ & $\blacksquare$
        & $\square$ & $\blacksquare$ & $\square$ \\

        \textbf{P9}
        & $\square$ & $\square$ & $\blacksquare$
        & $\blacksquare$ & $\blacksquare$ & $\blacksquare$
        & $\square$ & $\square$ & $\blacksquare$ \\

        \textbf{P12}
        & $\blacksquare$ & $\square$ & $\blacksquare$
        & $\blacksquare$ & $\square$ & $\square$
        & $\square$ & $\blacksquare$ & $\blacksquare$ \\

        \bottomrule
    \end{tabular}
\end{table}

Boundaries could vary substantially \textit{across participants for the same scenario}. To illustrate this variation, Table~\ref{tab:reasoning} compares three participants across three scenarios, using our thematic coding to indicate which dimensions were foregrounded in each participant's reasoning. For the example scenario of an LLM creating an image to accompany a post, participants invoked different considerations and weighed them differently. P12 foregrounded technology-related considerations, particularly the implications of image generation for artists and the use of their work in model training, leading them to avoid using GenAI for that purpose. P4 instead questioned whether generating an image was necessary in the first place, but reasoned that if someone ultimately chose to do so, they should at the very least disclose the use of GenAI. In contrast, P9's reasoning weighed considerations across all three dimensions: given the computational and energy costs they associated with generating images, they placed a higher bar on whether using GenAI served a meaningful purpose (rather than simply replacing an existing image), as well as whether its use was disclosed. We provide annotated excerpts from these participants' responses to this scenario in Appendix~\ref{app:annotated_excerpts}, offering a more detailed view of the reasoning summarized here.

\subsubsection{\textbf{The same participant drew different boundaries across scenarios}}
\label{sec:findings_boundaries_same_participants}

Similar variation also occurred \textit{within} participants across scenarios. A participant's considerations about GenAI did not necessarily disappear when they found another scenario appropriate, but changing what GenAI was being asked to do or the context in which it was used could change the weight of those considerations. P15, for example, viewed GenAI shortening an announcement positively, noting that in this case losing some information through ``lossy compression'' was not a problem. Yet, they drew a different boundary around using the same capability to shorten something in which a person had ``poured out [their] heart,'' reasoning that it's ``not going to communicate the same thing.'' Similarly, the salience of particular considerations could itself vary across scenarios. P9 foregrounded environmental concerns when considering GenAI generating an image, but less so when considering GenAI writing a post because they associated image generation with greater energy use.

\subsubsection{\textbf{Concrete scenarios surfaced and clarified participants' boundaries}}
\label{sec:findings_boundaries_eliciation}

Working through concrete scenarios sometimes raised new reflections that complicated participants' broader characterizations of their stance toward GenAI. P1, for example, described their default position as ``AI bad!'' but reflected that through the discussion, ``[their] viewpoint had definitely changed,'' as scenarios surfaced cases in which they felt GenAI could be useful. P10 similarly concluded that they were ``not as anti-AI as I might have thought.'' We also observed these reflexive, iterative reconsiderations while reasoning through a particular scenario during an interview session. After initially expressing concern that GenAI summarizing posts could separate creators from their audiences, P12 considered its potential for helping people understand the context behind recently trending posts or topics and reconsidered mid-thought: ``Okay, I've actually softened my take on this, because I can actually imagine that being kind of useful.'' Yet, engaging with concrete scenarios did not necessarily change participants' broader positions. P7 felt their views had not changed dramatically, but that they had ``clarified where [they] feel most comfortable,'' while P19 said their fundamental stance remained relatively the same but that the scenarios raised possibilities they hadn't thought of before. As such, considering GenAI through particular scenarios could surface, clarify, or complicate participants' boundaries even when their broader position toward GenAI remained unchanged.


\section{Discussion}
\label{sec:discussion}
Our findings identify a problem that is especially important for decentralized governance: participants' boundaries were not clean divisions between appropriate and inappropriate GenAI configurations nor stable positions for or against it. 
Instead, where participants drew their boundaries depended on the configuration of technology, integration, and use as well as which considerations became salient and how participants weighed them. These boundaries varied significantly across and within participants, reflecting a non-uniform relationship between considerations and judgments. Similar boundaries could arise from different considerations, while similar concerns could lead participants to draw different boundaries. The considerations underlying boundaries were often intertwined with what participants valued about decentralized social spaces, including being alternatives to centralized infrastructure, facilitating interaction and expression between people, and enabling greater agency over how shared spaces operate. 

Although DSM allows individuals to shape their spaces through collective decisions, our findings illustrate that decisions around GenAI may be difficult: the boundaries brought into governance are heterogeneous, conditional, and often only partially articulated. In what follows, we discuss how this boundary-based account complicates binary understandings of GenAI appropriateness in DSM (\S\ref{sec:discussion_drawing_boundaries}), discuss how such boundaries might be articulated individually (\S\ref{sec:discussion_boundary_articulation}) and negotiated collectively (\S\ref{sec:discussion_boundary_deliberation}), before considering how our findings may extend beyond DSM (\S\ref{sec:discussion_beyond_dsocial}).

\subsection{Understanding and Responding to GenAI Appropriateness Through Boundaries}
\label{sec:discussion_drawing_boundaries}

\subsubsection{\textbf{From binary labels to boundaries.}} 
Public discourse in DSM often frames stances toward GenAI as a dichotomy: one is either ``pro-AI'' or ``anti-AI''~\cite{MastodonAntiAI,MastodonLeftistAI,MastodonUnavoidableAI,MastodonStronglyAntiAI,BskyBacklashAI,BskyDisgustAI}. Although these labels can summarize overall sentiments about GenAI, our findings suggest that they obscure important nuances in how people reason about GenAI applications that can be important to collective decision-making. 
Our findings suggest that understanding GenAI appropriateness requires looking beyond the valence of a person's overall stance to the conditions underlying their judgments. We conceptualize this reasoning as a process of \emph{drawing boundaries}. A boundary is not a single threshold of acceptance, nor simply an observation that appropriateness depends on context. It consists of conditions that delimit which GenAI configurations a person considers appropriate across dimensions of \textit{technology}, \textit{integration}, and \textit{use}. This framing directs attention to what those conditions are and how people weigh them when judging particular GenAI configurations.

\subsubsection{\textbf{Characterizing boundaries across applications}} 
Our \textit{technology}, \textit{integration}, and \textit{use} dimensions provide a common vocabulary for characterizing the conditions underlying people’s reactions to GenAI in DSM. 
Prior work has shown that people's responses to GenAI can be conditional and situated~\cite{Schroeder25LLMQualitative,Kyi25GenAICreativeWork,Anthis26AISenseMaking}. Yet, such distinctions can be obscured when reactions are summarized as general attitudes toward GenAI or examined separately across particular applications. By distinguishing concerns about the underlying system, what GenAI is being asked to do, and how it is incorporated into a shared space, our dimensions make these heterogeneous judgments more comparable.

Our findings further suggest that even a nuanced understanding of a person's reaction in one context may not transfer to another, given that which considerations become salient and how they are weighed vary. In DSM, for example, support for GenAI in moderation~\cite{Zhang26Moderation} or rule compliance~\cite{LaCava25GenAIComplianceDsocial} may not necessarily indicate where people would draw their boundaries around GenAI for content creation, curation, or other uses. A person may accept the underlying technology but object to a particular use, or support a use only under specific integration conditions. Evidence that people support GenAI in one configuration should therefore not by itself be taken as support for introducing it elsewhere in the same ecosystem.

\subsubsection{\textbf{Interpreting and responding to acceptance and backlash.}} Beyond helping researchers characterize reactions more precisely, the three dimensions may also help developers and DSM community leaders interpret and respond to acceptance or backlash. As an example, we look back at Attie, the LLM-powered tool for creating custom Bluesky feeds using Anthropic's Claude that became the second-most-blocked account on Bluesky within days of its launch~\cite{BskyAttieSecondMostBlocked,BskyAttieUsesClaude}. Rather than interpreting such a response simply as anti-AI sentiment, our dimensions surface different considerations that could plausibly underlie it:
\begin{itemize}
    \item \textbf{Technology -- Who provides and controls the GenAI?} Objections might be about reliance on a model provided by a centralized AI company (like Anthropic) instead of infrastructure controlled locally or communally.

    \item \textbf{Integration -- How is the GenAI-powered feature introduced into the shared space?} Objections might be rooted in how a feature accesses and uses data from across the network, how those practices are communicated, and what control people retain over their data and participation.

    \item \textbf{Use -- Does the use of GenAI add meaningful value?} Objections might question whether an LLM meaningfully improves feed creation when custom feeds and other mechanisms for curating feeds already exist.
    \end{itemize}
These considerations do not necessarily explain any particular individual's decision to reject Attie. They instead highlight how the same outward response may reflect multiple underlying boundaries that call for different interpretations and responses, rather than a singular anti-AI position.

\subsection{Supporting Individual Boundary Articulation}
\label{sec:discussion_boundary_articulation}

\subsubsection{\textbf{Articulating boundaries may need further scaffolding}} 
As we conducted our interviews, we observed that participants did not always come in with their boundaries fully articulated. Instead, reasoning through concrete scenarios appeared to help some participants clarify their own thinking over the course of the interview session\footnote{For example, P10 reflected after discussing multiple scenarios that they felt they were not as anti-AI as they might have thought, while P7 described how the interview helped them clarify where they feel most comfortable.}. As such, nuanced boundaries may not always be readily captured when asking individuals for their positions toward GenAI. Articulating them can take time and prompting to consider concrete configurations, compare reactions across them, and reflect on which considerations matter or how GenAI should be involved in shared spaces.

Our observation that participants sometimes clarified or reconsidered their reasoning across scenarios also suggests value in examining how boundaries are articulated and potentially revised over time. Prior work has similarly cautioned against relying on static measurements of AI perceptions; for example, a review of 200 CHI and FAccT papers on perceptions of algorithmic fairness found that studies were predominantly ``cross-sectional'' and short in duration~\cite{VanBerkel23SnapshotsOfAttitudes}. Future work on GenAI in shared spaces could complement such measurements by eliciting reactions across multiple applications, revisiting boundaries as conditions change, or examining them longitudinally. For practitioners and community leaders, a corresponding implication is that governance around GenAI may similarly need to remain iterative, revisiting decisions as new technological properties, integrations, and uses are proposed. 

\subsubsection{\textbf{Leveraging participatory methods for boundary articulation}} 
One way to support boundary articulation is to draw from approaches such as participatory design, which offers a rich collection of methods for supporting this process, including scenarios, role-playing, mock-ups, and low-fidelity prototyping~\cite{Muller93ParticipatoryDesign,Svanaes04ParticipatoryDesignPrototyping,Druin99ParticipatoryDesignCooperativeInquiry}. Rather than using these methods only to elicit requirements, our findings suggest an opportunity to support people through a reflective process of examining whether and how a technology (such as GenAI) should be introduced into their shared spaces in the first place. This is particularly relevant in DSM, where decisions about the technologies that shape a space often involve users, developers, moderators, and administrators rather than being determined solely by a centralized platform. 

Our three dimensions offer one structure for constructing such activities. Participatory processes might, for example, present concrete scenarios that systematically vary the technical properties of GenAI, how it is integrated into a shared space, and what it does, and then ask individuals to reason about which changes make a scenario more or less appropriate to them. Comparing reactions across these scenarios could help individuals identify the considerations and conditions salient to their own boundary, producing more specific positions than a general stance of supporting or opposing GenAI. Future work might instantiate such activities through lightweight mechanisms such as \textit{boundary cards}, analogous in spirit to Model Cards~\cite{Mitchell19ModelCards}: structured artifacts that make otherwise implicit considerations explicit. Rather than documenting an ML model's properties, boundary cards could prompt individuals to consider alternative configurations across the three dimensions, record conditions that move a scenario within or outside their boundary, and reflect on which considerations matter most. 

\subsection{From Individual Boundaries to Collective Deliberation}
\label{sec:discussion_boundary_deliberation}

\subsubsection{\textbf{Structuring collective deliberation.}}
Articulating individual boundaries does not always tell us how we should respond to them collectively. Our findings highlight complexity around boundaries and that disagreement cannot always be reduced to opposing ``pro-AI'' and ``anti-AI'' camps. In DSM, where users, developers, moderators, and administrators may all help shape decisions about their shared spaces, these individual differences can intuitively make collective decision-making about how to introduce, manage, or deploy GenAI-related features challenging. 

Our findings offer one structure for making such disagreement more concrete. For example, a community revising its AI policy might elicit which conditions members care about across the technology, integration, and use dimensions and use these responses to identify points of convergence and disagreement. Existing social-computing approaches for deliberation and democratic decision-making~\cite{Kawakami24SituateAI,Kriplean12ConsiderIt,Fan20DigitalJurries,Yeo24HelpMeReflect} offer mechanisms that could then be adapted for deliberating boundaries. The \textit{Situate AI Guidebook}, for instance, scaffolds multi-stakeholder deliberation about proposed AI systems through structured questions spanning different aspects of a deployment~\cite{Kawakami24SituateAI}; our dimensions could similarly provide prompts through which members articulate and compare the conditions underlying their boundaries. Systems such as \textit{ConsiderIt} could help make disagreements legible by encouraging participants to articulate tradeoffs and engage with others' reasoning~\cite{Kriplean12ConsiderIt}, while approaches such as \textit{Digital Juries} suggest how smaller groups might deliberate over concrete cases when substantial disagreement remains~\cite{Fan20DigitalJurries}.

\subsubsection{\textbf{Deliberation need not produce uniformity.}} Importantly, collective deliberation does not need to culminate in a single boundary shared by every individual within the shared space. P19, for instance, suggested a ``consent model'' through which people could signal whether they were comfortable with practices such as having their posts ingested by AI systems, alongside mechanisms for interacting with others whose preferences were compatible with their own. Other circumstances may instead require communities to establish shared minimum conditions, prohibit applications that conflict with strongly held community values, or provide individual-level controls that allow multiple boundaries to coexist. With this in mind, an important direction for future work is to examine which forms of collective decision-making are appropriate when boundaries converge, partially overlap, or remain fundamentally incompatible. Surfacing boundaries on its own will not resolve disagreement or produce consensus, but it can make the conditions underlying disagreement visible enough to deliberate over.

\subsubsection{\textbf{From boundaries to action.}} Our findings also raise a related question: once a boundary condition has been identified, who has the capacity to act on it? Different conditions may implicate different actors and layers of the sociotechnical system within DSM. Some conditions rooted in the technology dimension may require action from model developers or providers; conditions concerning integration may be more actionable by platform or application developers as well as administrators; and conditions related to use may be more directly shaped through community norms and expectations around how members use GenAI. These mappings are neither exclusive nor fixed. A privacy concern, for example, may implicate both the model developer and the application developer. Nonetheless, locating where a condition can be acted upon can help distinguish disagreements that may be addressed locally through design or policy from those that require engagement with infrastructures or actors beyond a shared space's immediate control.

\subsection{Boundary Drawing Beyond Decentralized Social Media}
\label{sec:discussion_beyond_dsocial}

Our findings are grounded in Mastodon and Bluesky, and the particular considerations participants foregrounded should be interpreted in light of this setting. Decentralization, autonomy, and agency are key to these online spaces and likely shaped why participants cared about where control was located, whether GenAI infrastructure was centralized, who could opt out, and whether its integration aligned with local norms. The \textit{technology}, \textit{integration}, and \textit{use} dimensions therefore emerged in a setting where questions about authority, control, and infrastructure are particularly salient.

At the same time, boundary drawing may provide a useful lens for understanding judgments about GenAI beyond DSM. Other shared spaces, such as centralized social platforms or collaboratively governed projects, also make decisions about the appropriateness of GenAI. In such settings, people may similarly reason about characteristics of the technology, how it is integrated, and what it is used to do.
More broadly, this lens could help us study how people reason about the appropriateness of GenAI in their everyday lives. A person need not hold a single position toward GenAI that transfers across, for example, creative work, professional tasks, and interpersonal communication. 
Our findings do not establish that the same dimensions or considerations structure judgments similarly in these contexts but motivate future work examining whether boundary drawing provides a useful structure for understanding how people distinguish appropriate from inappropriate applications of GenAI across domains.

\section{Limitations}
Our findings are shaped by our participants' subjectivity and self-presentation, as well as the interpretive nature of our qualitative analysis. Although we strove to interview individuals with varying backgrounds and roles, our participant pool of 20 represents a relatively small subset of Mastodon and Bluesky individuals. Our findings should not be interpreted as a definitive overview of what matters to people in DSM regarding GenAI. For example, GenAI and DSM are both rapidly evolving. Changes in both entities may alter which considerations and dimensions are salient and where people draw their boundaries, highlighting the need to study boundary articulation as an ongoing governance process through more longitudinal work. 

Further, our findings capture boundaries elicited through scenario-based interviews rather than those established in practice. The scenarios we presented helped participants surface and refine their reasoning but may have also shaped which considerations became salient. In particular, our scenarios primarily varied how GenAI was used, rather than systematically varying factors related to technology or integration, given that the importance of the latter two became salient to us over the course of this study. This initial design may have contributed to the relative prevalence of use-related considerations in our interview data. Accordingly, we do not interpret the frequency with which dimensions or considerations arose as indicative of their relative importance. Instead, we emphasize that the three dimensions of technology, integration, and use along with the considerations within them, offer a structure for making discussions about GenAI appropriateness more nuanced and informative for supporting collective deliberation within DSM.

\section{Conclusion}
As GenAI enters DSM, the core governance challenge is not simply whether people support or oppose it, but determining whether and under what conditions particular applications of GenAI are considered appropriate. Our interview study shows that people draw conditional boundaries around GenAI based on considerations about the technology, its integration, and its use, with these boundaries varying across people and within the same person across scenarios. This heterogeneity matters especially in DSM, where authority is distributed across multiple actors and decentralization creates opportunities for them to shape norms and decisions in shared spaces. With this in mind, governing GenAI in DSM may benefit from supporting people in articulating their boundaries and deliberating across areas of agreement and disagreement. Framing GenAI appropriateness as a process of drawing boundaries offers a way to move beyond simple pro- or anti-AI positions common in public discourse in DSM toward more situated and actionable decisions about GenAI in decentralized social spaces.

\section*{Generative AI Disclosure}
We used GenAI tools to enhance the search for related works and refine the writing and formatting of this manuscript. Specifically, we used Claude, ChatGPT, and Elicit to find relevant research papers for both the related works and discussion sections (alongside non-GenAI tools, like Google Scholar). After the Discussion was written, we used ChatGPT to refine and streamline the wording of the written content, which we manually verified and edited again. We also used Claude for specific formatting tasks, such as generating table formats. Where GenAI has been used for editing, we certify that we have read, adapted, and corrected the text as necessary, and stand behind the resulting work.

\bibliographystyle{ACM-Reference-Format}
\bibliography{references}

\appendix
\section{Interview Protocol}
\label{app:interview_protocol}
We provide the interview protocol used for our semi-structured interviews below. Note that the term ``operators'' refers to individuals who have been involved in moderation efforts or administering instances. 

\noindent \textbf{Background}. Participants were first asked about their general experiences with DSM and GenAI.
\begin{itemize}
    \item Which decentralized social media platforms do you use?
    \item Can you tell me the story of how you joined these platforms?
    \begin{itemize}
        \item How do you use these platforms?
        \item How have you liked it so far?
    \end{itemize}
    \item Have you ever used GenAI in any way in / for decentralized social media?
    \begin{itemize}
        \item If so, can you give a specific example?
        \begin{itemize}
            \item What was that experience like?
        \end{itemize}
        \item If not, why not?
    \end{itemize}
    \item \textbf{[Developers / Operators]} Have you worked on any projects or tools for decentralized social media?
    \begin{itemize}
        \item As part of your work on [project / tool], did you / your team ever use GenAI?
        \item If so, how was GenAI used?
        \begin{itemize}
            \item Was GenAI used as part of the development / moderation / administration process?
            \item Was GenAI used as part of the project / tool itself?
        \end{itemize}
        \item Why was GenAI used / not used?
    \end{itemize}
\end{itemize}

\noindent \textbf{Mental Models of AI and GenAI}. Participants were next asked to describe their understanding of ``AI'' and ``GenAI'' before discussing specific scenarios, with the goal of grounding the conversation in a shared understanding of GenAI and knowing what technological objects they considered as such.

\begin{itemize}
    \item How would you describe ``AI'' to someone else?
    \item How would you describe ``GenAI'' to someone else?
    \item Do you see ``AI'' and ``GenAI'' as the same thing or different things?
    \item When you hear the term ``AI,'' what are the top three words that come to mind?
    \item When you hear the term ``GenAI,'' what are the top three words that come to mind?
\end{itemize}

\noindent \textbf{Scenarios}. Participants were then presented with seven scenarios about GenAI in decentralized social media. For each scenario, participants were asked about their thoughts, reactions, and the factors shaping their reactions. 
\begin{itemize}
    \item \textbf{Scenario 1 -- Text Generation}: An LLM used to generate a post from user-written bullet points.
    \item \textbf{Scenario 2 -- Image Generation}: An LLM used to generate an accompanying image for a user-written post.
    \item \textbf{Scenario 3 -- Length Adjustment}: An LLM used to shorten a user-written post to fit a character limit.
    \item \textbf{Scenario 4 -- Writing Revision}: An LLM used to revise the grammar and spelling of a user-written post.
    \item \textbf{Scenario 5 -- Summarization}: An LLM used to summarize posts in a feed or timeline.
    \item \textbf{Scenario 6 -- Curation}: An LLM used to curate a feed or timeline based on a user's stated preferences.
    \item \textbf{Scenario 7 -- Articulation}: An LLM used to help users articulate their preferences for what they want to see in a feed or timeline.
\end{itemize}

Questions asked for each scenario:
\begin{itemize}
    \item How would you feel about this scenario?
    \begin{itemize}
        \item Why? What makes you feel this way?
        \item Would you personally use GenAI in this way?
        \item Would you be okay with others using GenAI in this way?
        \item Is there anything that, if done or changed, would make you feel very differently?
    \end{itemize}
    \item Have you already seen or experienced a scenario like this?
\end{itemize}

\noindent \textbf{Reflections}. Participants were asked to reflect more broadly on the future of GenAI in DSM.
\begin{itemize}
    \item Thinking about the future of decentralized social media, what role (if any) would you want GenAI to have in this space?
    \begin{itemize}
        \item Are there certain possibilities you'd be excited about?
        \item Are there certain possibilities you'd want avoided?
    \end{itemize}
    \item Is there anything you wish users, developers, moderators, or admins understood or kept in mind about GenAI in the context of decentralized social media?
    \item Thinking back over everything we've discussed today, have any of your views changed or become clearer in any way?
    \item Is there anything we didn't discuss that you think is important?
\end{itemize}

\section{Codebook}
\label{app:codebook}
We provide the shared codebook developed during thematic analysis in Table~\ref{tab:codebook}. Codes added by the first author after the initial codebook development are highlighted in gray and marked with asterisks (***). These additions capture (1) concerns about opacity and bias in GenAI systems and (2) expectations that GenAI features keep users' interests at the forefront without conflicting or ulterior motives. Codes whose definitions were subsequently expanded by the first author are also highlighted in gray. These revisions include (1) broadening the human involvement code to encompass users taking ownership of the final output; (2) broadening the suitability code to capture both whether GenAI is appropriate for a task and whether it can perform that task effectively; and (3) broadening the testing and iteration code to encompass evaluation both before and after deployment.

\newpage
\begingroup
\footnotesize
\begin{longtable}{p{0.33\textwidth} | p{0.6\textwidth}}
\caption{\textbf{The shared codebook developed during our thematic analysis}. Codes revised after the initial development of the shared codebook (3 codes) are highlighted in gray; codes added subsequently (2 codes) are highlighted in gray and marked with asterisks (***).}
\Description{Two-column table presenting our shared thematic analysis codebook. Each row contains a code and its description. Codes whose definitions were revised after the initial codebook development are highlighted in gray. Codes added after the initial codebook development are highlighted in gray and additionally marked with three asterisks.}
\label{tab:codebook} \\

\toprule
\textbf{Code} & \textbf{Description} \\
\midrule
\endfirsthead

\toprule
\textbf{Code} & \textbf{Description} \\
\midrule
\endhead

\bottomrule
\endfoot

\textbf{Vague Mental Models / Definitions of GenAI}
& How participants define, distinguish, or blur AI and GenAI, including uncertainty about what makes a technology ``generative.'' \\

\textbf{Vagueness Attributed to Marketing / Big Tech}
& Attributions of ambiguity around AI or GenAI terminology to marketing, technology companies, or strategic use of the ``AI'' label. \\

\textbf{Participant Examples of GenAI in DSM}
& Examples of GenAI use cases or scenarios participants have encountered, used, or imagined in decentralized social media. \\

\textbf{Conditionality of GenAI Acceptance}
& Statements around how the appropriateness of GenAI depends on the context or nature of its use as well as statements about participants realizing nuance in their own positions. \\

\hline

\textbf{Resource \& Environmental Costs}
& Energy, water, carbon, monetary, or other resource costs associated with GenAI, including concerns about wastefulness or efficiency. \\

\rowcolor{gray!10}
\textbf{Opacity \& Bias in GenAI***}
& Concerns about the opacity, black-box nature, or biases of GenAI systems and the difficulty of understanding how they operate or produce outputs. \\

\textbf{Training Data Practices \& Privacy}
& Concerns about how data are collected, used, or obtained for model training, including consent, copyright, ownership, scraping, and privacy. \\

\textbf{Corporate Influence \& Big Tech Distrust}
& Concerns about centralization, corporate control, business incentives, data collection, or distrust of large technology companies providing GenAI systems. \\

\textbf{Broader Sociotechnical Impacts}
& Broader consequences of GenAI, such as labor displacement, effects on learning, dependency, reduced human skill or effort, and other societal impacts. \\

\textbf{Alternative Solutions \& Workarounds}
& Alternatives proposed to mitigate GenAI-related concerns, such as local or smaller models, conventional algorithms, or non-GenAI tools. \\

\hline

\textbf{Address Real User \& Community Needs}
& Expectations that GenAI systems or features address meaningful needs of users or communities rather than introducing technology for its own sake. \\

\rowcolor{gray!10}
\textbf{Maintain Users at Forefront***}
& Expectations that GenAI systems or features prioritize users' interests, including protecting privacy and avoiding conflicting or ulterior motives. \\

\rowcolor{gray!10}
\textbf{Test Quality \& Iterate Accordingly}
& Expectations that GenAI systems or features be thoroughly tested, evaluated, and iteratively improved before and during deployment. \\

\textbf{Transparency Around GenAI}
& Expectations for disclosure of GenAI involvement and transparency about how a GenAI system or feature operates. \\

\textbf{Optionality \& Agency Over GenAI}
& Expectations that users retain meaningful choice and control over GenAI features, including the ability to opt in, opt out, or customize their use. \\

\hline

\textbf{Neutral Reactions to Offloading Uses}
& Neutral reactions to GenAI taking over substantive generation or creation with limited human involvement. \\

\textbf{Negative Reactions to Offloading Uses}
& Negative reactions to GenAI taking over substantive generation or creation with limited human involvement. \\

\textbf{Positive Reactions to Assistive Uses}
& Positive reactions to GenAI supporting a human-led task. \\

\textbf{Neutral Reactions to Assistive Uses}
& Neutral reactions to GenAI supporting a human-led task. \\

\textbf{Preserve Human Authenticity}
& Concerns about voice, self-expression, genuineness, originality, or whether content continues to represent the person producing it. \\

\textbf{Limit Input-Output Transformation}
& Whether GenAI preserves the user's intended meaning, ideas, or goals, including the degree of transformation between input and output. \\

\textbf{Ensure Quality \& Reliability}
& Concerns about accuracy, hallucinations, errors, ``AI slop,'' output quality, or difficulty assessing GenAI-generated output. \\

\rowcolor{gray!10}
\textbf{Maintain Human Involvement}
& Expectations that users remain involved, understand limitations, verify outputs, contribute effort, and take responsibility for the final result. \\

\rowcolor{gray!10}
\textbf{Verify Suitability of GenAI for Task}
& Whether GenAI is an appropriate, necessary, useful, or efficient tool for a task, including whether better non-GenAI alternatives exist. \\

\end{longtable}
\endgroup

\section{Annotated Excerpts for the Image-Generation Scenario}
\label{app:annotated_excerpts}

The excerpts below provide a more detailed view of how three participants reasoned about the image generation scenario discussed in \S\ref{sec:findings_boundaries_diff_participants}. Highlighted passages indicate reasoning associated with the dimensions of technology in \techhl{blue}, integration in \integrationhl{purple}, and use in \usehl{green}, with our annotations identifying the corresponding considerations. 

\paragraph{\textbf{P4 --- Integration + Use.}}
``I would ask, \usehl{can you not search Google images and the things that are open source, and find one that way?}
\textit{\textbf{[Use: Verify suitability of GenAI for task].}} [\ldots] I mean, I think that \usehl{you can be specific in your prompting and get maybe something that has quality, iterate by asking it to take out the third arm} \textit{\textbf{[Use: Maintain human involvement].}} [\ldots] But I get that it's going to happen. It is happening, and it's going to continue to happen. And \integrationhl{I do think that it should be
disclosed}
\textit{\textbf{[Integration: Transparency around GenAI].}} [\ldots] It's like for people who are skeptics, like maybe me, or people who are even more skeptical than I am, then it's good to, \integrationhl{I really think it should be disclosed for image generation, even for text generation, I think you should say, we used AI in this way} \textit{\textbf{[Integration: Transparency around GenAI].}}'' 

\paragraph{\textbf{P9 --- Technology + Integration + Use.}}
``I think a lot of people use AI to augment their own work, but not replace them, and \usehl{images are one of the places where it's very clear that AI was used in replacement of either finding an image by another person,
or generating it yourself} \textit{\textbf{[Use: Maintain human involvement].}} \techhl{It's also a very intensive form of AI usage, where it consumes significantly more resources} \textit{\textbf{[Technology: Resource and environmental costs].}} [\ldots] \integrationhl{I think I would just want it to be clearly labeled. There are services on Bluesky that I think attempt to label AI images, so some
way of identifying that it is an AI-generated image and not real} \textit{\textbf{[Integration: Transparency around GenAI].}}''

\paragraph{\textbf{P12 --- Technology.}}

``I think it's even more I think pronounced for visual
artists, the idea that people who have no ability to create art, create visual art, suddenly have the ability to just type what they want, and then get something resembling visual art out of it. Like, I can totally get how
that'd be super frustrating. \techhl{I honestly really avoid doing generative AI art just purely for that reason, because I do feel that artists are probably gonna be the most impacted by things of that nature, so I try not to use it for that purpose} \textit{\textbf{[Technology: Training data practices].}}''

\end{document}